\documentclass[aps,prd,reprint,superscriptaddress,nofootinbib]{revtex4-2}
\usepackage{graphicx}
\usepackage{amsmath,amssymb}
\usepackage{hyperref}
\hypersetup{
    citecolor=blue,
    colorlinks=true,
    linkcolor=blue,
    filecolor=magenta,      
    urlcolor=blue,
    pdftitle={Overleaf Example},
    pdfpagemode=FullScreen,
    }

\newcommand\apjl{ApJ}                % Astrophysical Journal, Letters
\newcommand\araa{ARA\&A}             % Annual Review of Astronomy and Astrophysics
\newcommand\jcap{J.~Cosmology Astropart. Phys.} % Journal of Cosmology and Astroparticle Physics
\newcommand\aap{A\&A}                % Astronomy and Astrophysics
\newcommand\aplett{Astrophys.~Lett.} % Astrophysics Letters

\newcommand{\EQ}[1] {Eq.~(\ref{#1})}

\newcommand{\SEC}[1] {Sec.~\ref{#1}}
\newcommand{\APP}[1] {Appendix~\ref{#1}}

\usepackage{xcolor}

\definecolor{dkblue}{RGB}{54, 86, 169}

\begin{document}

%\preprint{}

\title{Identifying axion conversion in compact star magnetospheres with radio-wave polarization signatures}


\author{Z.~H.~Xue}
\affiliation{Department of Astronomy, Peking University, Beijing 100871, China}
\affiliation{National Astronomical Observatories, Chinese Academy of Sciences, 20A Datun Road, Chaoyang District, Beijing 100101, China}

\author{K.~J.~Lee}
\email{kjlee@pku.edu.cn}
\affiliation{Department of Astronomy, Peking University, Beijing 100871, China}
\affiliation{National Astronomical Observatories, Chinese Academy of Sciences, 20A Datun Road, Chaoyang District, Beijing 100101, China}
\affiliation{Kavli Institute for Astronomy and Astrophysics, Peking University, Beijing 100871, China}

\author{X.~D.~Gao}
\email{gaoxiangdong@bjut.edu.cn}
\affiliation{Faculty of Science, Beijing University of Technology, Beijing 100124, China}

\author{R.~X.~Xu}
\affiliation{Department of Astronomy, Peking University, Beijing 100871, China}
\affiliation{State Key Laboratory of Nuclear Physics and Technology, School of Physics, Peking University, Beijing 100871, People’s Republic of China}

%\thanks{}
% \affiliation{}


\date{\today}

\begin{abstract}
    The axion is well motivated in physics. It solves the strong charge
conjugation-parity reversal problem \textit{CP} in fundamental physics and the dark matter problem in astronomy. Its interaction with the electromagnetic field has been expected but never detected experimentally. Such particles may convert to radio waves in the environment with a strong magnetic field. Inspired by the idea, various research groups have been working on theoretical modeling and radio data analysis to search for the signature of radio signals generated by the axion conversion in the magnetosphere of compact stars, where the surface magnetic field as strong as $10^{13}$--$10^{14}$ G is expected.
    In this work, we calculate the observational properties of the axion-induced radio signals (AIRSs) in the neutron star magnetosphere, where both the total intensity and polarization properties of radio emission are derived. Based on the ray tracing method, assuming 100\% linear polarization of radio waves generated in each conversion,  we compute the polarization emission profile concerning different viewing angles. We note that plasma and general relativistic effects are important for the polarization properties of AIRSs. Our work suggests that AIRSs can be identified by the narrow bandwidth and distinct polarization features.
\end{abstract}

%\keywords{}

\maketitle

\section{Introduction}\label{sec:intro}
Since the 1970s, the Standard Model of particle physics has been the most successful description of all fundamental interactions other than gravity. It is a quantum field theory that has in many cases been verified experimentally mainly with accelerators. Despite the successes, the Standard Model has some theoretical shortcomings. One of these issues is the strong \textit{CP} problem. The strong \textit{CP} problem is a puzzle that the observations (e.g., the vanishing small electric dipole moment for neutrons) indicate a conserved \textit{CP} in spite of the fact that the theory needs fine-tuning to meet such a symmetry.

Dark matter constitutes the vast majority ($\sim 30\%$) of matter in our Universe. The evidence for the existence of dark matter has been abundant and convincing over the past several decades (e.g., \cite{1933AcHPh...6..110Z,1970ApJ...159..379R,1998ApJ...498L.107T,2021ARA&A..59..247H}).  Nevertheless, it seems that no particles within the Standard Model can play the role required of being cold, stable, and weakly coupled.

Axions may solve the strong \textit{CP} problem \citep{1977PhRvL..38.1440P} and dark matter problem \cite{1983PhLB..120..127P,1983PhLB..120..133A,1983PhLB..120..137D} in a coherent way. Axions are expected to have the following properties. First, they are pseudoscalar particles. Therefore, the axion’s equation of motion is described by the simple Klein-Gordon equation. Second, the axion mass $m_a$, induced by quantum chromodynamics (QCD) instantons can be calculated in chiral perturbation theory \cite{1978PhRvL..40..223W}, which is given by $m_{a,{\rm QCD}}\approx 6\times 10^{-6}\ {\rm eV}\left(10^{12}\ {\rm GeV}/(f_a/\mathcal{C})\right)$, where $f_a/\mathcal{C}$ is a model-dependent value. Third, it is expected to weakly interact with the electromagnetic field $\mathcal{L}\propto g_{a\gamma\gamma}a\mathbf{E}\cdot\mathbf{B}$.

At present, a large number of terrestrial experiments have been made to probe the parameter space of axion-photon couplings. The axion haloscopes have been designed to search for photon signals from axions using the microwave cavity resonators. Experiments based on such ideas include examples of the Rochester–Brookhaven–Fermilab \citep{PhysRevLett.59.839}, University of Florida \citep{1990PhRvD..42.1297H} and ADMX \citep{2010PhRvL.104d1301A,2018PhRvL.120o1301D,2020PhRvL.124j1303B,PhysRevLett.127.261803}.

Despite those terrestrial experiments, a variety of astrophysical and cosmological observations can also be used to constrain the parameter space of axions or axionlike particles. The ratio of horizontal branch stars to red giants in the Galactic globular clusters is affected by axion-photon conversion inside stars and can be used to place a constraint on axion-photon couplings \citep{2014PhRvL.113s1302A}. A cosmological example is using the x-ray emission excess from axion-photon conversion in the magnetic fields offered by the clusters of galaxies \citep{2013PhRvL.111o1301C}.

Recently, it is widely discussed that axions may convert to radio frequency photons in the magnetosphere of neutron stars, a compact star with surface strong magnetic field \citep{mt77}. The radio signature of such a process is a narrow spectral line at a frequency mainly determined by the axion mass \citep{2009JETP..108..384P,2018PhRvL.121x1102H,2020PhRvD.102b3504B,2021JCAP...11..013M}. Reference \citep{2020Leroy} first employed the ray tracing method to obtain detailed observational properties of such process. This was further developed to account for light refraction and Doppler broadening in \citep{richard2021,2021PhRvD.104j3030W}. Also, a variety of groups have begun to work on analyzing radio data to search for evidence of axion conversion \citep{2020PhRvL.125l1103D,2020ApJ...900L..28D,2020PhRvL.125q1301F,2022PhRvD.105b1305B,2022PhRvL.129y1102F}.

In this work, we investigate the polarization properties of axion conversion in the magnetosphere of compact stars. Polarizations have been briefly discussed in the supplemental material of \citep{2018PhRvL.121x1102H}, where they gave an analytical description of the polarization profile by assuming radial trajectories for axions and converted photons. We extend the up-to-date ray tracing method to provide more accurate polarization properties by building a detailed numerical model for calculating axion-induced emission from compact star systems. Those unique polarization signatures can be helpful in the future to identify axion-induced radio signals (AIRSs) and mitigate potential false alarms. 

The remainder of this paper is organized as follows. In \SEC{sec:methods}, we describe the methods for calculating axion-induced emission, paying particular attention to polarization properties. In \SEC{sec:results}, we show the results of our calculation. Finally, we give a conclusion in \SEC{sec:conclusion}.

\section{Methods}\label{sec:methods}

\subsection{Axion-photon conversion}\label{subsec:axion-photon-conv}
To derive the axion-photon conversion probability, we start from the standard Lagrangian for the axion, photon, and their couplings. Using the Euler-Lagrangian equation, Ohm's law, and assuming that the dielectric tensor and magnetic field are time invariant, the equations of motion for electric field $\mathbf{E}$ and axion field $a$ are (see, e.g., \citep{2018PhRvL.121x1102H,2020PhRvD.102b3504B,2021PhRvD.104j3030W,2021JCAP...11..013M})
\begin{align}
    -\partial_t^2a +\nabla^2a &= m_a^2 a-g_{a\gamma\gamma}{\mathbf{E}}\cdot {\mathbf{B}}\, , \label{eq:axion-EOM} \\
    -\nabla^2 {\mathbf{E}}+\nabla(\nabla\cdot {\mathbf{E}}) &= \omega^2 {\mathbf{D}}+\omega^2 g_{a\gamma\gamma}a{\mathbf{B}}\, , \label{eq:E-EOM}
\end{align}
where $\mathbf{D}=\mathbf{\epsilon E} $ is the electric displacement field and $\mathbf{\epsilon}$ is the dielectric tensor. Although the plasma of the pulsar magnetosphere is believed to be relativistic \citep{RS75}, we take the magnetoionic approximation (magnetized subrelativistic plasma) in the current paper and leave the work of relativistic plasma modeling for future work. Under the magnetoionic approximation, the dielectric tensor depends on plasma frequency and gyrofrequency (see \APP{app:dielectric-tensor}). Then, in the short planar wave limit, the radio wave and axion interaction is described by \citep{2018PhRvL.121x1102H,2021PhRvD.104j3030W}
\begin{equation}
    -\partial_z^2\left(\begin{matrix}E_y \\ a\end{matrix}\right) = \left(\begin{matrix}\frac{\omega^2-\omega_p^2}{1-\frac{\omega_p^2}{\omega^2}\cos^2\tilde{\theta}} & \frac{g_{a\gamma\gamma}B_t\omega^2}{1-\frac{\omega_p^2}{\omega^2}\cos^2\tilde{\theta}} \\ \frac{g_{a\gamma\gamma}B_t}{1-\frac{\omega_p^2}{\omega^2}\cos^2\tilde{\theta}} & \omega^2-m_a^2\end{matrix}\right)\cdot\left(\begin{matrix}E_y \\ a\end{matrix}\right), \label{eq:eom-ey-a}
\end{equation}
where $\omega$ and $\omega_p$ are the photon and plasma frequency (angular frequency), $\tilde{\theta}$ is the angle between magnetic field and direction of axion propagation, and $B_t$ is the component of the magnetic field transverse to the axion propagation. Taking the Wentzel-Kramers-Brillouin (WKB) approximation will help to simplify the first equation to a Schr\"{o}dinger-like equation, which has a solution in integration form. Using stationary phase approximation, this solution can be further simplified to give the axion-photon conversion probability (\APP{app:axion-electrodynamics}), 
\begin{equation}
    P_{a\to\gamma}=\frac{1}{\omega^2}\left|\frac{E_y}{a}\right|^2=\frac{\pi g_{a\gamma\gamma}^2B_t^2}{2\omega_p'(\mathbf{x}_c)v_c}\, , \label{eq:convprob}
\end{equation}
where $\omega_p'(\mathbf{x}_c)$ and $v_c$ are the derivative of plasma frequency along the axion trajectory and axion velocity at the point of resonant conversion, which is given by the stationary point condition,
\begin{equation}
    \omega_p(\mathbf{x}_c)^2 = \frac{m_a^2 \omega^2}{m_a^2\cos^2\tilde{\theta}+\omega^2\sin^2\tilde{\theta}}\, . \label{eq:conv-condi}
\end{equation}
In the nonrelativistic limit, the energy-momentum relation becomes $\omega\to m_a$, and the resonance condition simplifies to $\omega_p(\mathbf{x}_c)=m_a$. 
Note that Eq.~(\ref{eq:convprob}) diverges when $\omega_p'\to 0$. This occurs near the edge of images, which was pointed out and discussed in \citep{richard2021}. The divergence is caused by the breakdown of WKB approximation. One needs to solve \EQ{eq:eom-sch} for the rigorous results or use a conservative remedy like truncating the conversion length by some typical value \citep{2021JCAP...11..013M}. Here, we take the approach of Ref.~\citep{richard2021} by simply excluding the parameter space where WKB approximation is invalid.

\subsection{Magnetosphere}\label{subsec:magnetosphere}
In order to determine the plasma frequency, a detailed model describing the magnetosphere of compact stars is required. We adopt the classical Goldreich-Julian (GJ) model \citep{1969ApJ...157..869G}. The GJ model assumes a static and corotating magnetosphere in the presence of a dipole magnetic field given by\footnote{Strictly speaking, the GJ model is not applicable to inclined rotator nor active pulsars \citep{1982RvMP...54....1M}. However, since we are mainly interested in the region close to the star surface, the GJ model provides a reasonable start.}
\begin{align}
    B_r &= B_0\left(\frac{R}{r}\right)^3\left(\cos\chi\cos\theta+\sin\chi\sin\theta\cos\psi\right)\, , \label{eq:dipolefield1}    \\
    B_\theta &= \frac{B_0}{2}\left(\frac{R}{r}\right)^3\left(\cos\chi\sin\theta-\sin\chi\cos\theta\cos\psi\right)\, , \label{eq:dipolefield2}   \\
    B_\phi &= \frac{B_0}{2}\left(\frac{R}{r}\right)^3\sin\chi\sin\psi\, , \label{eq:dipolefield3}
\end{align}
where $\chi$ is the inclination angle, $\psi = \phi-\Omega t$, $B_0$ is the surface magnetic field of a compact star, $R$ is the radius, and $\Omega = 2\pi/P$ is the angular rotation velocity. The GJ model gives the charge number density,
\begin{equation}
    n_{\rm GJ}(\mathbf{r}) = \frac{2\mathbf{\Omega}\cdot\mathbf{B}}{e}\frac{1}{1-\Omega^2r^2\sin^2\theta}\, .
\end{equation}
Assuming $n_c = |n_{\rm GJ}|$ and neglecting the relativistic corrections in the denominator, the conversion surface can be derived using
\begin{equation}
    \omega_p = \sqrt{\frac{4\pi e^2|n_{\rm GJ}|}{m_e}},
\end{equation}
and the resonant conversion condition $\omega_p=m_a$. In practice, the GJ density may not give the true charge density in the magnetosphere; that electron density can be 100--1000 times larger \citep{RS75}. In addition, dipole magnetic field configuration may not hold when approaching the surface of compact stars; it will be more appropriate to include multipolar fields \citep{GMM02}. Although we tested the effects of multipolar fields with a perturbation of approximately 10\% quadrupole components, more rigorous work is left for future work.

\subsection{Axion density}\label{subsec:DMdensity}
In this paper, as axions are regarded as the dark matter counterpart, the terms ``axion'' and ``dark matter'' are interchangeable in the following text. In general, the intensity of AIRSs equals axion flux density times axion-photon conversion probability. Thus, the distribution function is required to derive the local axion density near a compact star (CS). According to Liouville's theorem, the distribution function is conserved along trajectories, namely,
\begin{equation}
    f(\mathbf{r}_{\rm CS},\mathbf{v}) = f_\infty(\mathbf{r}_\infty,\mathbf{v}_\infty)\, ,
\end{equation}
where $f_\infty(\mathbf{r}_\infty,\mathbf{v}_\infty)$ denotes the distribution function far away from the compact star, and $\mathbf{v}_\infty = \mathbf{v}_\infty(\mathbf{r}_{\rm CS},\mathbf{v})$ is the velocity far away from the star for an orbit with velocity $\mathbf{v}$ at location $\mathbf{r}_{\rm CS}$. 
Assuming that the velocity distribution of dark matter (in the Galactic rest frame) is given by the Maxwell-Boltzmann distribution and is isotropic (in the frame of a compact star) far away from the compact star, then the distribution function is given by
\begin{equation}
    f_\infty(\mathbf{r}_\infty,\mathbf{v}_\infty) = \frac{\rho_\infty^{\rm DM}}{(\pi v_0^2)^{3/2}}\exp\left(-\frac{\mathbf{v}_\infty^2}{v_0^2}\right)\, ,\label{eq:df_MB}
\end{equation}
where $\rho_\infty^{\rm DM}$ is the local Galactic dark matter density far away from the compact star and $v_0$ is the velocity dispersion of dark matter.  Because of the conservation of energy, the relation between $\mathbf{v}_\infty$ and $\mathbf{v}$ is \citep{2006PhRvD..74h3518A,2020Leroy}
\begin{equation}
    \mathbf{v}_\infty^2 = \mathbf{v}(r)^2 - \frac{2GM}{r}\, ,
\end{equation}
where $M$ is the mass of the compact star. Substituting this into Eq.~(\ref{eq:df_MB}) and integrating over all allowed velocities, $|\mathbf{v}| \ge \sqrt{2GM/r}$, leads to the dark matter density at a given radius $r$. Furthermore, since gravity dominates near the star surface, i.e.,  $2GM/r\gg v_0^2$, the enhanced local dark matter density can be further approximated by \citep{2006PhRvD..74h3518A,2018PhRvL.121x1102H,2020Leroy,2021JCAP...11..013M}
\begin{equation}
    \rho(r) \simeq \frac{2\rho_\infty^{\rm DM}}{\sqrt{\pi}}\sqrt{\frac{2GM}{r}}\frac{1}{v_0}\, . \label{eq:dm-density}
\end{equation}
Here, the enhancement of dark matter density is a natural consequence of compressing the momentum space volume.
Note that the assumption that dark matter velocity distribution is isotropic in the frame of a compact star (i.e., $\mathbf{v}_{\rm CS}=0$) is not necessarily valid, since the compact star and dark matter are both moving subrelativistically, where their velocity is of the same order of magnitude \citep{2020Leroy,2021JCAP...11..013M}. At present, we use the isotropic dark matter density (\ref{eq:dm-density}) for further calculation and leave the anisotropic case for future investigations.

\subsection{Polarization}\label{subsec:polarization}
Polarization, a statistical property of quasimonochromatic electromagnetic waves, is described by the Stokes parameters. Following \citep{1979rpa..book.....R}, for electric field components $E_1$ along $\hat{\mathbf{x}}$ and $E_2$ along $\hat{\mathbf{y}}$, the Stokes parameters ($\{I,Q,U,V\}$) are defined via the second-order correlation of the electric field,
\begin{align}
    I & \equiv  \langle E_1 E_1^\ast\rangle + \langle E_2 E_2^\ast\rangle\, ,  \label{eq:si} \\
    Q & \equiv  \langle E_1 E_1^\ast\rangle - \langle E_2 E_2^\ast\rangle\, , \label{eq:sq} \\
    U & \equiv  \langle E_1 E_2^\ast\rangle + \langle E_2 E_1^\ast\rangle\, ,  \label{eq:su} \\
    V & \equiv  -i\left(\langle E_1 E_2^\ast\rangle - \langle E_2 E_1^\ast\rangle\right)\, . \label{eq:sv}
\end{align}
The $x$-$y$ plane is defined perpendicular to the direction of the converted photon's trajectory. Without loss of generality, $\hat{\mathbf{x}}$ is chosen to be coplanar with the direction of the compact star's spin rotation $\hat{\mathbf{\Omega}}$; in other words,
\begin{equation}
    \hat{\mathbf{y}} = \hat{\mathbf{\Omega}} \times \hat{\mathbf{k}}\, ,\quad
    \hat{\mathbf{x}} = \hat{\mathbf{k}} \times \hat{\mathbf{y}}\, , \label{eq:basis}
\end{equation} 
where $\hat{\mathbf{k}}$ is the unit vector of the photon's 3-momentum. After determining the direction of momentum, we can calculate the Stokes parameters in terms of the above bases.

We now show how to calculate polarization within our framework using Eqs.~(\ref{eq:si})--(\ref{eq:sv}). Polarization properties can be inferred from the first equation of Eqs.~(\ref{eq:eom-ey-a}), where
\begin{equation}
    \frac{g_{a\gamma\gamma}B_t\omega^2}{1-\frac{\omega_p^2}{\omega^2}\cos^2\tilde{\theta}}a\, ,
\end{equation}
acts as a source term to the electric field $E_y$. Hence, this axion-induced electric field is aligned with the tangential component of the compact star's magnetic fields $B_t$. Suppose that the magnetic field is $\mathbf{B}$ at the conversion point, using the bases defined in Eqs.~(\ref{eq:basis}), the angle between $\mathbf{B}_t$ and $\hat{\mathbf{x}}$ is given by
\begin{equation}
    \cos\theta = \frac{\mathbf{B}\cdot\hat{\mathbf{x}}}{\sqrt{\left(\mathbf{B}\cdot\hat{\mathbf{x}}\right)^2+\left(\mathbf{B}\cdot\hat{\mathbf{y}}\right)^2}}\, ,
\end{equation}
with the electric field denoted $\mathbf{E}$ whose magnitude is proportional to the square root of the intensity of axion-induced emission (namely, $E\propto\sqrt{I}$), defining
\begin{equation}
    \mathcal{E}_1 = E\cos\theta, \quad \mathcal{E}_2 = E\sin\theta\, . \label{eq:e1-e2}
\end{equation}
Thus, the Stokes parameters are calculated as
\begin{align}
    I_i &= \mathcal{E}_{1,i}^2 + \mathcal{E}_{2,i}^2\, , \\
    Q_i &= \mathcal{E}_{1,i}^2 - \mathcal{E}_{2,i}^2\, , \\
    U_i &= 2 \mathcal{E}_{1,i} \mathcal{E}_{2,i}\, , \\
    V_i &= 0\, ,\label{eq:v_i}
\end{align}
for arbitrary photon $i$. Here, $V_i$ is set to zero so that the polarization is 100\% linearly polarized for each converted photon, since $\mathbf{E}$ is perfectly aligned with $\mathbf{B}_t$. As the AIRSs are incoherent emissions, the total Stokes parameters are
\begin{align}
    I &= \sum_i I_i\, , \label{eq:tot-I} \\
    Q &= \sum_i Q_i\, , \\
    U &= \sum_i U_i\, , \\
    V &= \sum_i V_i\, . \label{eq:tot-V}
\end{align}
With the total Stokes parameters, one can derive degree of linear polarization ($\Pi$) and position angle ($\chi$) using the definitions of
\begin{equation}
    \Pi = \frac{\sqrt{Q^2+U^2}}{I}\, ,\quad \chi = \frac{1}{2}\arctan\frac{U}{Q}\, . \label{eq:pa-pi}
\end{equation}

\subsection{Ray tracing}\label{subsec:raytracing}
Ray tracing is a technique of computer graphics to simulate lighting scenes. 
The ray tracing method obtains an image by tracing the path of the light originating from receivers toward the light sources. 
The basic algorithm of the ray tracing is as follows: (1) grid the image plane into small pixels; (2) determine each ray path according to the receiver position and the center of a pixel in the image plane; (3) trace each ray backward according to the geometric optics until hitting the emitter; and (4) sum the contribution of image pixels to compute the observed quantities.

This method was first introduced in Ref.~\citep{2020Leroy} to obtain the observational properties of axion-induced emission. It was further developed to consider an inhomogeneous and time-dependent magnetosphere as well as including the effects of gravity in \citep{richard2021}. At the same time, Ref.~\citep{2021PhRvD.104j3030W} also tried to investigate the plasma effects on the converted photons' propagation. However, Ref.~\citep{2021PhRvD.104j3030W} employed a forward ray tracing method, where photons start from the conversion point and are traced forward to the celestial sphere at infinity. The distinction between backward and forward ray tracing is only methodological and should not give a significant difference in the final results. In our framework, we use the backward ray tracing method as was used in \citep{2020Leroy,richard2021}.

With the ray tracing method, the flux of axion-induced emission can be calculated with \citep{2020Leroy,richard2021}
\begin{equation}
    F_{\rm obs} = \frac{(\Delta b)^2}{D^2}\sum_i\underbrace{\left(\frac{n_a v_a}{4\pi}P_{a\to\gamma}\right)(\theta_{\rm obs}^i,\phi_{\rm obs}^i)}_{\equiv I_{\rm obs}^i(\theta_{\rm obs}^i,\phi_{\rm obs}^i)}\, , \label{eq:flux}
\end{equation}
where $\Delta b$ is the size of each pixel, $D$ is the distance from the observer to the compact star, and $n_a$ and $v_a$ are the number density and velocity of axion dark matter. Our ray tracing procedures are concluded as follows.
\begin{enumerate}
    \item First, we define the projection image plane as a square plane perpendicular to the line of sight toward the compact star at a distance $D$ from Earth. This plane is located at a distance $d$ from the compact star. The latter distance is chosen to be larger than the maximum distance at which the resonant axion-photon conversion occurs. The total area of the projection image plane should be large enough to cover the whole conversion region. This plane is divided into square pixels of size $\Delta b$, whose center identifies a specific photon trajectory $i$.
    \item For each pixel, we simulate photon trajectory toward a compact star by numerically computing its geodesic starting from the center of the corresponding pixel. The direction of initial velocity is taken to be perpendicular to the projection plane.
    \item With a specific starting time for each photon, the plasma frequency can be calculated along each trajectory. The resonant conversion point is determined by solving equation $\omega_p(\mathbf{r}_c)=m_a$. The position $\mathbf{r}_c$ should be identified as the first point that comes to satisfy the resonant condition along each trajectory. Other possible solutions should be omitted since these photons must have traveled through a plasma region with $\omega_p>\omega$, therefore they are scattered during their travel. We do not need to worry about such cases if we include plasma effects since reflection is automatically embodied in the geodesics. Also, the simulation of photon propagation should be truncated once it hits the surface of the compact star since the resonant conversion should take place outside the compact star.
    \item The power of axion-induced emission from each pixel is computed in the conversion region using Eq.~(\ref{eq:flux}) [terms in the summation notation times the area of the pixel $(\Delta b)^2$]. Those pixels that do not have a crossing between photon trajectories and conversion surface of course do not contribute to the total emission. The polarization of each pixel is obtained using Eqs.~(\ref{eq:e1-e2})--(\ref{eq:v_i}) after the calculation of its corresponding power.
    \item The total axion-induced emission power is obtained by summing over all pixels in the projection image plane. The total Stokes parameters are calculated in the same way as is shown in Eqs.~(\ref{eq:tot-I})--(\ref{eq:tot-V}).
\end{enumerate}
We consider two photon propagation models in this paper, propagation in vacuum \citep{2020Leroy} and in cold plasma \citep{richard2021}. 
We start with deriving the covariant theory of photon propagation using the eikonal equation \citep{1962PhRv..126.1899W}, where the eikonal can be written as
\begin{equation}
    \Phi(\lambda_1, \lambda_2) = \int_{\lambda_1}^{\lambda_2}d\lambda\, k_{\mu}\frac{dx^\mu}{d\lambda}\, ,
\end{equation}
using calculus of variation and one obtains
\begin{equation}
    \frac{dk_\mu}{dx^\nu}\frac{dx^\mu}{d\lambda}-\frac{dk_\nu}{d\lambda} = 0\, . \label{eq:variation}
\end{equation}
Let the dispersion relation be written in the invariant form $D(k,x)=0$ in a slowly varying medium. The dispersion relation must be satisfied everywhere along the ray path; this requires not only $D(k,x)=0$ but also $d D(k,x)/d\lambda=0$ \citep{Plasmadynamics-Melrose}, i.e.,
\begin{equation}
    \frac{dD(k,x)}{d\lambda} = \frac{dk_\mu}{d\lambda}\frac{\partial D(k,x)}{\partial k_\mu}+\frac{dx^\mu}{d\lambda}\frac{\partial D(k,x)}{\partial x^\mu} = 0\, , \label{eq:dD=0}
\end{equation}
be satisfied along the ray path. Identifying Eqs.~(\ref{eq:variation}) and (\ref{eq:dD=0}) leads to
\begin{equation}
    \frac{dx^\mu}{d\lambda} = \frac{\partial D(k,x)}{\partial k_\mu}\, , \quad \frac{dk_\mu}{d\lambda} = -\frac{\partial D(k,x)}{\partial x^\mu}\, , \label{eq:ray-eq}
\end{equation}
which are the Hamiltonian equations \citep{1999prop.book.....B} for light propagation.
For a cold isotropic plasma, the dispersion relation is given by $D=g_{\mu\nu}k^\mu k^\nu-\omega_p^2$. Inserting this into the above equations gives
\begin{equation}
    \frac{d^2x^\mu}{d\lambda^2}+\Gamma_{\alpha\beta}^\mu\frac{dx^\alpha}{d\lambda}\frac{dx^\beta}{d\lambda} = -\frac{1}{2}g^{\mu\nu}\partial_\nu \omega_p^2\, , \label{eq:geodesic}
\end{equation}
where $\lambda$ is the worldline parameter. This is the equation used for computing the photon propagation in the presence of plasma in curved spacetime. Also, we can use Eqs.~(\ref{eq:ray-eq}) to eliminate $\lambda$ and just use coordinate time $t$ as the worldline parameter.

Including plasma refraction and curved spacetime has two accompanying effects: plasma lensing and gravitational redshift. They can be concluded in (see \citep{richard2021})
\begin{equation}
    \frac{I_{\rm obs}}{n_{\rm obs}^2\omega_{\rm obs}^3} = \frac{I_{\rm em}}{n_{\rm em}^2\omega_{\rm em}^3}\, ,
\end{equation}
where subscripts ``obs'' and ``em'' represent quantities obtained at the location of the projection image plane and the conversion region, respectively. Taking a Schwarzschild metric for the compact star gives gravitational redshift as
\begin{equation}
    \frac{\omega_{\rm obs}^3}{\omega_{\rm em}^3} = \left(1-\frac{r_s}{r_c}\right)^{3/2}\, ,
\end{equation}
where $r_s=2GM$ is the Schwarzschild radius of the compact star. Plasma lensing is included in the refractive index $n$, taking $n_{\rm obs}=1$, and \citep{richard2021}
\begin{equation}
    n_{\rm em}^2\approx \frac{1}{1-r_s/r_c}\left(\frac{r_s}{r_c}+v_0^2\right)\, ,
\end{equation}
thus, $n_{\rm obs}^2/n_{\rm em}^2=1/n_{\rm em}^2\gg 1$; this is where plasma lensing comes to work.

Furthermore, the influence on polarization due to light bending should be studied carefully. For curved photon trajectories, the directions of 3-momentum for different photons are different at the location of conversion points. Therefore, the basis vectors defined in \EQ{eq:basis} are not the same for different photons, and then Stokes parameters cannot be added directly using Eqs.~(\ref{eq:tot-I})--(\ref{eq:tot-V}). Instead, we utilize the following equation to parallel transport the polarization vectors \citep{2017mcp..book.....T}:
\begin{equation}
    \frac{d\hat{\mathbf{f}}}{ds} = -\hat{\mathbf{k}}\left(\hat{\mathbf{f}}\cdot\frac{d\hat{\mathbf{k}}}{ds}\right)\, , \label{eq:propagation-of-polari}
\end{equation}
where $\hat{\mathbf{k}}$ is the unit vector for the propagation direction, $s$ is the distance along the light ray, and $\hat{\mathbf{f}}$ is the unit polarization vector. This propagation law can be derived from Maxwell's equations using eikonal approximation and is shown to be consistent with Fermi-Walker transport in three-dimensional space. Also, if we regard the refractive index as a space metric, this formula implies that the unit polarization vector is parallel transported along the light ray (see \APP{app:polarization}). With this formula, unit polarization vectors can be transported to the projection image plane where the basis vectors are the same, whereupon we can add them incoherently as described in the ray tracing procedure.

At this stage, the initial conditions can be set for simulating the photon trajectories. The equation to be solved is Eq.~(\ref{eq:geodesic}), which consists of four second-order differential equations. We can divide them into eight first-order differential equations for numerical integration. Since we can eliminate the parameter $\lambda$ and use the coordinate time as the worldline parameter, we can just solve seven first-order differential equations. The initial time for each photon is set by the specific phase of the rotation period; the initial 3-coordinate is set by the location in the projection image plane. The 4-momentum comprises the frequency and 3-momentum. The initial frequency should satisfy $|\omega-m_a|/m_a\leq v_0^2\sim 10^{-6}$. The initial direction for the 3-momentum is set as described in the ray tracing procedures which is perpendicular to the projection image plane\footnote{Note that this is valid as long as the projection image plane is placed at a relatively large distance.}; the amplitude of the 3-momentum is determined by the frequency and dispersion relation. With the initial conditions all set up, we can numerically solve the geodesics and follow the procedures described above to obtain the axion-induced emission. Things get much easier for the vacuum case, where photons travel in straight lines. We just sample some points along a straight line and look for the local maxima of plasma frequency. With these local maxima of $\omega_p$, we can then search for the first point that satisfies $\omega_p=m_a$. This method can be much more efficient and accurate than propagating photons in a straight line.

Our code is written in \texttt{C++}. The fifth-order Runge-Kutta-Fehlberg method \citep{fehlberg1970classical} is employed to solve the coupled differential equations, where the initial integration step size is refined to achieve the relative error specification of $10^{-8}$.  Because of the geometric self-similarity, we use the radial distance as the scale of initial trial integration step size, so that $h = r/10^4$ with $r$ being the photons' altitude. During the step refining processes, the true step size can only be smaller than the initial value. For the plasma case, our code performs $5\times 10^5$ trajectory simulations within 3.5 h on a single CPU core, while it is much faster for the vacuum case, which can simulate $5\times 10^5$ trajectories within only 12 s. Nevertheless, with the help of importance sampling, the time cost of generating one projection image for the plasma case reduces to 10--20 min without loss of accuracy.

\section{Results}\label{sec:results}
As an example, based on the above discussions, we compute the AIRSs from a single isolated slow rotating neutron star, J0806.4-4123. Its rotation period is $P = 11.37\ {\rm s}$. It is a nearby system, i.e. $D\approx 250\ {\rm pc}$ from Earth \citep{2009ApJ...705..798K}. Its magnetic field is expected to be strong [surface magnetic field $B_0 = 2.5\times 10^{13}\ {\rm G}\,(1\ {\rm G}=10^{-4}\ {\rm T})$]. Those properties make it a good target to search for AIRSs.
We take the mass of the star to be $1\,M_\odot$, the radius of $10\ {\rm km}$, and $\rho_{\rm DM}^\infty = 0.3\ {\rm GeV/cm^3}$ and $v_0 = 200\ {\rm km/s}$ for the local density and velocity dispersion for dark matter particles, respectively.

\subsection{Projection image}\label{subsec:proj-image}
\begin{figure}
    \centering
    \includegraphics[width=0.9\hsize]{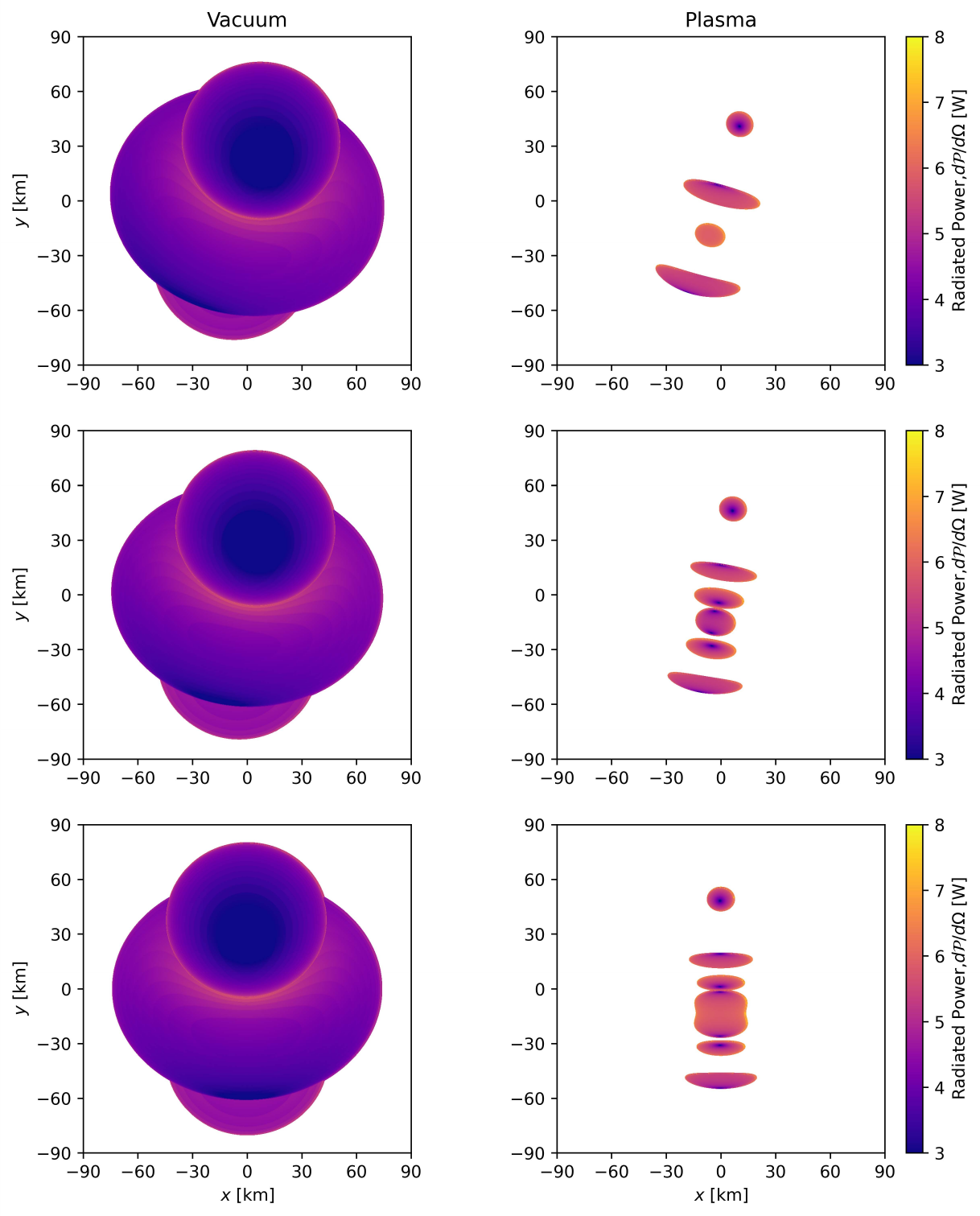}
    \caption{Radiated power in projection image plane: vacuum versus plasma. Snapshots of the radio photons at frequencies $\omega = m_a$ produced by axion conversion in the magnetosphere as seen in the projection plane perpendicular to a viewing angle of $\theta=36^\circ$. The benchmark scenario is chosen with an inclination angle of $\chi=18^\circ$, $g_{a\gamma\gamma}=10^{-12}\ {\rm GeV}^{-1}$, and $m_a=0.5\ {\rm \mu eV}$. The left column shows the results of straight line propagation through vacuum, reproducing the results of \citep{2020Leroy}. The right column results from including plasma effects, reproducing the results of \citep{richard2021}. From top to bottom, we show the results for pulse phases of 0.3, 0.4, and 0.5. All six plots share the same color bar which is shown next to the right column and they are shown in log scale.}
    \label{fig:proj-image}
\end{figure}

The intensity of AIRSs in the projection image plane is shown in Fig.~\ref{fig:proj-image}. The viewing angle (the angle between line of sight and the spin axis) of these projection images is $\theta=36^\circ$ with an inclination angle of $\chi=18^\circ$. The axion mass chosen for calculation is $m_a=0.5\ {\rm \mu eV}$ and the axion-photon coupling constant is $g_{a\gamma\gamma}=10^{-12}\ {\rm GeV}^{-1}$. From the figure, one can see that including plasma and general relativistic effects can cause significant differences in the AIRSs. As is also mentioned in \citep{richard2021}, the radiation region gets smaller and splits up when including plasma effects. Although the image size is shrunk relative to the vacuum case, the total axion-induced emission power does not decrease significantly (see the colors in Fig.~\ref{fig:proj-image} or the axion-induced emission pulse profile shown later), in some cases, it even \emph{increases} significantly. Note that our calculations of projection image for the vacuum case reproduce the results derived in \citep{2020Leroy} and for the plasma case reproduce the results obtained in \citep{richard2021}.

\subsection{Axion-induced emission pulse profile}\label{subsec:profile}
\begin{figure*}
    \centering
    \includegraphics[width=0.9\hsize]{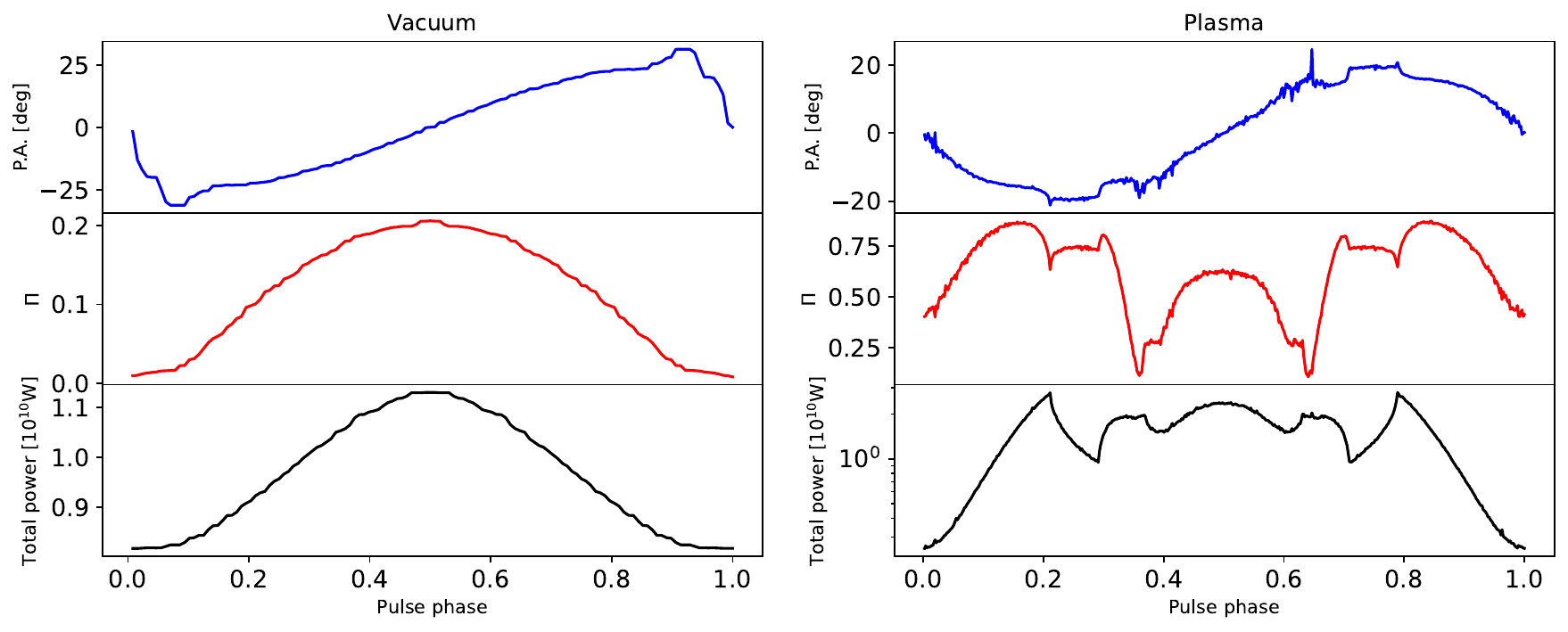}
    \caption{Axion-induced radio emission as a function of pulse phases: vacuum versus plasma. A comparison of axion-induced emission pulse profiles and polarizations without and with plasma is shown in the left and right columns, respectively. Each column shows the position angle (P.A.) of polarization (upper), the degree of linear polarization [see Eq.~(\ref{eq:pa-pi})] (middle), and the total power (unit in watts) of axion-induced emission (lower). The viewing angle of this figure is $36^\circ$. Other parameters are set the same as the benchmark scenario used in Fig.~\ref{fig:proj-image}.}
    \label{fig:profile-36}
\end{figure*}

\begin{figure*}
    \centering
    \includegraphics[width=0.9\hsize]{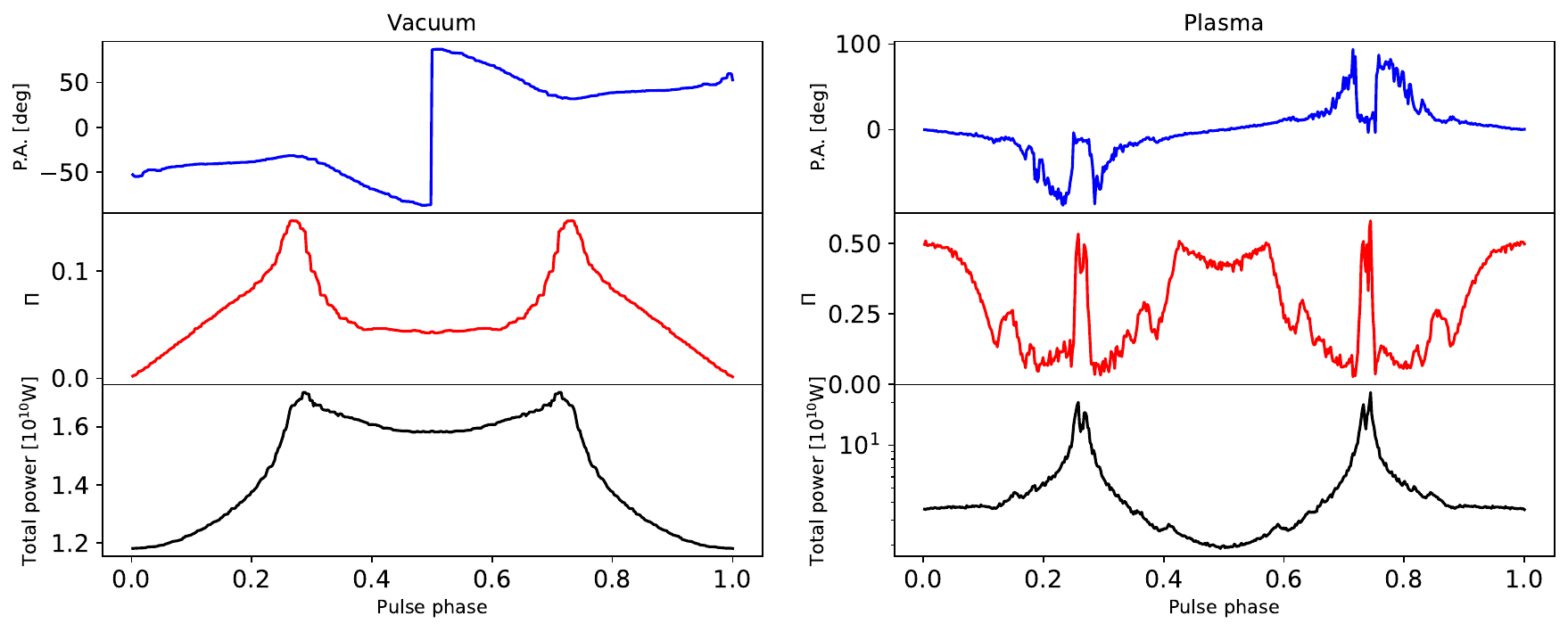}
    \caption{Axion-induced radio emission as a function of pulse phases: vacuum versus plasma. This figure is the same as Fig.~\ref{fig:profile-36}, only with a different viewing angle of $54^\circ$.}
    \label{fig:profile-54}
\end{figure*}

\begin{figure*}
    \centering
    \includegraphics[width=0.9\hsize]{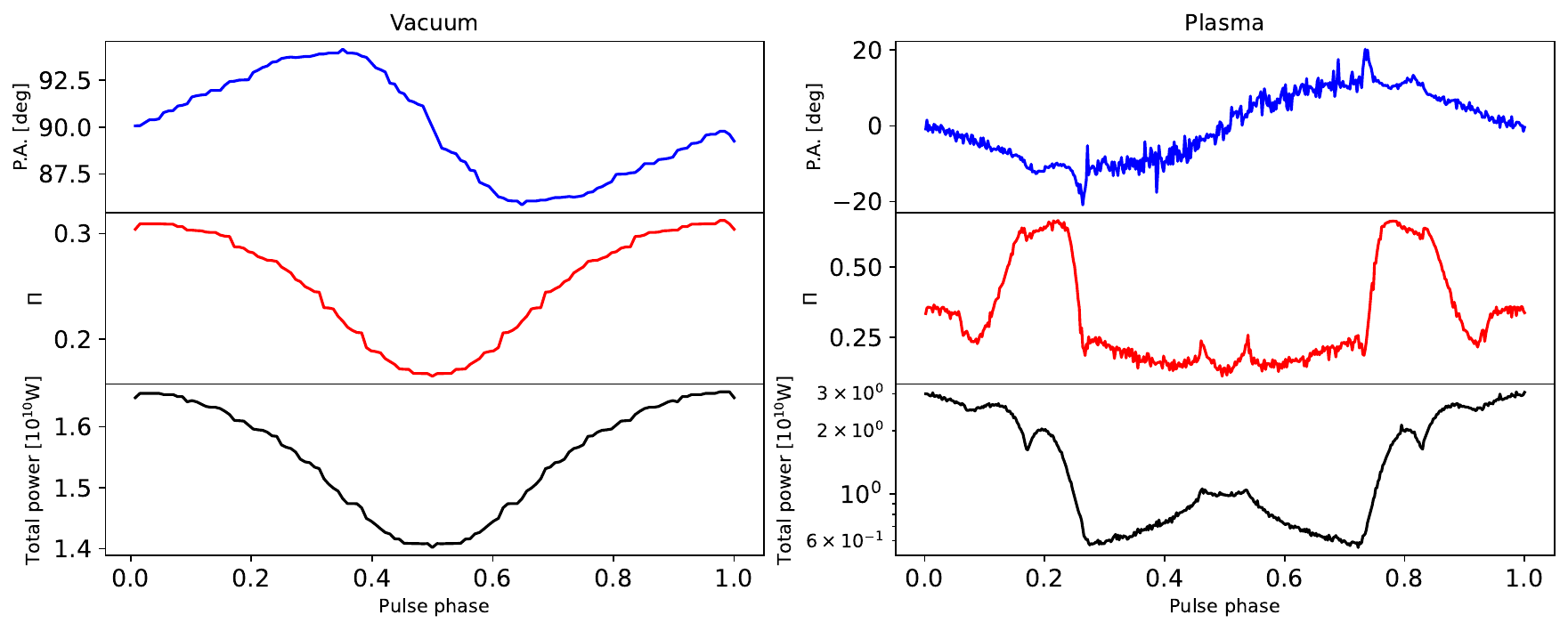}
    \caption{Axion-induced radio emission as a function of pulse phases: vacuum versus plasma. This figure is the same as Fig.~\ref{fig:profile-36}, only with a different viewing angle of $72^\circ$.}
    \label{fig:profile-72}
\end{figure*}

In Figs.~\ref{fig:profile-36}--\ref{fig:profile-72} we report the total axion-induced emission power (assuming the contribution of all the pixels) along with polarization properties as a function of pulsar rotational phases. The same as the projection image, we show the results of both the vacuum case and plasma case for comparison. The three figures correspond to three different viewing angles $\theta=36^\circ,\,54^\circ$, and $72^\circ$, while inclination angle, axion mass, and axion-photon coupling constants are the same as before. Within each figure, the upper panel is the position angle of polarization, the middle panel is the degree of linear polarization, and the lower one is axion-induced emission pulse profile in units of watts.

From the three lower panels, as mentioned above, including plasma effects does not decrease the maximum radiated power significantly. However, the profile shape changed significantly so that the contrast (i.e., the pulse amplitude divided by mean flux) of the signal pulse becomes higher. It also causes larger variations concerning different viewing angles. From the three middle panels, a larger variation also occurs in polarizations. First, we find that the degree of linear polarization changes along with axion-induced emission power for the vacuum case, while it does not evolve along with total power in the plasma case and gets more complicated. Second, the degree of linear polarization in the vacuum case ranges from 0 to 0.2 ($\theta=36^\circ$), while it gets higher in the plasma case, ranging from 0.1 to 0.8 ($\theta=36^\circ$). This can be explained qualitatively. In our framework of calculating polarizations, we assumed 100\% linear polarization for each pixel. To obtain total polarizations, we sum over all pixels incoherently, which decreases the degree of linear polarization. However, as can be seen from Fig.~\ref{fig:proj-image}, including plasma effects splits the projection image into several smaller regions, and within each region the degree of linear polarization is high due to the pixels therein sharing similar position angles. Thus, summing over these localized highly linearly polarized regions produces a higher degree of linear polarization.

\subsection{Convergence and robustness tests}\label{subsec:convergcheck}
We check the convergence and robustness of our calculation by comparing the results computed with different setups, i.e., number of pixels, integration step size, and small perturbations to magnetic field models. The results show that our calculation achieves convergence at the $7\%$ level for different numbers of pixels and $3\%$ level for a different choice of step size. Also, we note that a 10\% perturbation to the dipole magnetic field configuration will not make a significant difference to the AIRS pulse profile or polarization. The details can be found in \APP{app:convergence}.

\section{Discussions and Conclusions}\label{sec:conclusion}
In this paper, we have developed the framework to numerically compute the AIRSs from the magnetosphere of neutron star systems, where the intensity and polarization properties can be computed with the ray tracing method.

We find that an inclined magnetic field configuration of a neutron star leads to nontrivial polarization structures in AIRSs. Our calculation reveals that the AIRSs are dominated with linear polarization, which can be even higher considering plasma effects in photon propagation. We note that the position angle of linear polarization for the AIRSs does not follow the well-known ``S'' curve seen in many pulsars and modeled by the rotating vector model  \citep{1969ApL.....3..225R}. In our case, although the direction of the electric field of AIRSs at the emission point still follows the local magnetic field direction, the parallel transport and superposition of polarization vectors lead to a more complex situation. Thus, the AIRSs can be identified from pulsar emission (if any) by (1) a narrow band radio signal with a frequency centered at axion mass, (2) a distinct polarization feature within the narrow frequency band compared to the neighbor frequencies, and (3) linear polarization being dominant in the narrow band.  Given that radio frequency interference is usually circularly polarized and pulsar radiation properties in general evolve smoothly as functions of frequency, the polarization feature will help us truly ``identify'' the axion events in future observations.

There are three caveats in our modeling. We excluded the region where the WKB approximation is invalid, assumed an isotropic distribution of dark matter particles, and adopted the stationary GJ magnetosphere model. As one would expect, the AIRS flux is low in the region where WKB approximation is invalid and dark matter should be isotropic at the small scale of the pulsar magnetosphere; the key issue here is the unknown magnetic field configuration.
We had estimated the effects of the magnetic field configuration via tests of perturbing the dipole field. Since the dominant radiation originates from the low altitude (strong magnetic field and higher plasma density), we expect the surface magnetic field of neutron stars plays a vital role in determining the properties of AIRSs. In the future, a more realistic model on AIRSs is required to account for the effect.

We compare our results with those in \citep{2020Leroy,richard2021}. Our results for the vacuum case agree perfectly with \citep{2020Leroy}, whereas for the plasma case, there exist some differences with \citep{richard2021}. The main difference comes from the scenario with a viewing angle of $\theta=54^\circ$ and axion mass of $m_a=0.5\ {\rm \mu eV}$. We find that the profile shows two local maxima (or peaks) at the pulse phase of 0.25 and 0.75. The same conflict remains when we switch off the plasma effects. Peaks at such phases are expected. In the GJ magnetosphere model, there exists a null charge surface where plasma frequency is zero (i.e., where $\mathbf{\Omega}\cdot\mathbf{B}=0$). This surface is located at a polar angle of $\theta=\arccos(1/\sqrt{3})\approx54^\circ$. 
Therefore, photons at low altitudes can propagate along the null charge surface slit without being significantly affected by the plasma effects. Those photons are at the pulse phase of 0.25 and 0.75 with an inclination angle of $18^\circ$ and a viewing angle of $54^\circ$. A more recent work \citep{2023arXiv230311792B} resembles the double-peak features of our results at phase 0.25 and 0.75. However, due to different settings of model parameters, it is hard to evaluate if the results are compatible with our findings.

At low altitudes, the axion-photon conversion probability gets higher due to the higher magnetic field Eq.~(\ref{eq:convprob}), thus giving a peak to the pulse profile at those phases. Thus, if the viewing angle is $36^\circ$, then we would expect a peak at the pulse phase of 0.5 (since $36^\circ=-18^\circ+54^\circ$), as is shown in Fig.~\ref{fig:profile-54}. If the viewing angle is $72^\circ$, then a peak should be expected at the pulse phase of 0 or 1 (since $72^\circ=18^\circ+54^\circ$, see Fig.~\ref{fig:profile-72}).

\begin{acknowledgments}
This work is supported by the National SKA Program of China (2020SKA0120100), the National Key R\&D Program of China (2017YFA0402602), the National Nature Science Foundation Grant No. 12041303, the CAS-MPG LEGACY project, and funding from the Max-Planck Partner Group.
\end{acknowledgments}

\appendix

\section{Dielectric tensor} \label{app:dielectric-tensor}
In this appendix, we describe the derivation of the cold plasma dielectric tensor relevant for solving axion electrodynamics in Sec.~\ref{subsec:axion-photon-conv}. We follow the single particle theory here. For the current problem, the single particle theory delivers the same results as the more refined kinetic theory for plasma \citep{1989plwa.book.....S}. The equation of motion for a nonrelativistic charged particle is
\begin{equation}
    m\frac{\mathrm{d}\mathbf{v}}{\mathrm{d}t} = q(\mathbf{E}+\mathbf{v}\times\mathbf{B})\, , \label{eq:eom-particle}
\end{equation}
along with the expression for the current,
\begin{equation}
    \mathbf{j} = nq\mathbf{v}\, .
\end{equation}
For the time-harmonic field with a static background magnetic field $B_0$, we have
\begin{align}
    \mathbf{E} &= \mathbf{E}_1 e^{i(\mathbf{k}\cdot\mathbf{r}-\omega t)}\, , \\
    \mathbf{B} &= \mathbf{B}_0+\mathbf{B}_1 e^{i(\mathbf{k}\cdot\mathbf{r}-\omega t)}\, , \\
    \mathbf{v} &= \mathbf{v}_1 e^{i(\mathbf{k}\cdot\mathbf{r}-\omega t)}\, ,
\end{align}
where the coordinate is chosen such that static magnetic field $\mathbf{B}_0$ is along the $z$ direction, and $|\mathbf{B}_0 | \gg |\mathbf{B}_1|$. Hence, \EQ{eq:eom-particle} becomes
\begin{equation}
    -i\omega m \mathbf{v}_1 = q\left(\mathbf{E}_1+\mathbf{v}_1\times\mathbf{B}_0\right)\, ,
\end{equation}
where the second-order terms (e.g., $\mathbf{v}_1\times\mathbf{B}_1$) have been neglected. The solution to this equation is given by 
\begin{align}
    v_x &= \frac{iq}{m\left(\omega^2-\omega_c^2\right)}\left(\omega E_x+i\,\mathrm{sgn}(q)\omega_c E_y\right)\, ,\\
    v_y &= \frac{iq}{m\left(\omega^2-\omega_c^2\right)}\left(-i\,\mathrm{sgn}(q)\omega_c E_x+\omega E_y\right)\, ,\\
    v_z &= \frac{iq}{m\omega}E_z\, ,
\end{align}
where $\mathrm{sgn}(q) = q/|q|$ denotes the sign of the charge and $\omega_c = |q|B_0/m$ is the cyclotron frequency. Using Ohm's law, $\mathbf{j}=\boldsymbol{\sigma}\cdot\mathbf{E}$, and the definition for dielectric tensor $\boldsymbol{\epsilon}=\mathbb{I}-4\pi\frac{\boldsymbol{\sigma}}{i\omega}$, the electric displacement field is 
\begin{equation}
    \mathbf{D} = R^{yz}_{\theta}\cdot\left(\begin{matrix}\varepsilon & ig & 0 \\ 
    -ig & \varepsilon & 0 \\ 
    0 & 0 & \eta\end{matrix}\right)\cdot R^{yz}_{-\theta}\cdot\left(\begin{matrix}E_x \\ E_y\\ E_z\end{matrix}\right)\, , \label{eq:electric-displacement}
\end{equation}
where the external magnetic field $\mathbf{B}$ lies in the $(y,z)$ plane at an angle $\theta$ with positive $z$ axis. $R^{yz}_{\theta}$ is the rotation matrix in the $(y,z)$ plane. The coefficients $\varepsilon, g$, and $\eta$ are
\begin{equation}
    \varepsilon = 1-\frac{\omega_p^2}{\omega^2-\omega_c^2}\, , \, g = \frac{\omega_p^2\omega_c}{\omega(\omega^2-\omega_c^2)}\, , \, \eta = 1-\frac{\omega_p^2}{\omega^2}\, .
\end{equation}
For high frequency limit ($\omega_c\gg \omega,\omega_p$), $\varepsilon\to 1$ and $g\to 0$, and the displacement field becomes
\begin{equation}
    \mathbf{D} \simeq \left(\begin{matrix}
        1 & 0 & 0 \\
        0 & 1-\frac{\omega_p^2}{\omega^2}\sin^2\theta & \frac{\omega_p^2}{\omega^2}\cos\theta\sin\theta \\
        0 & \frac{\omega_p^2}{\omega^2}\cos\theta\sin\theta & 1-\frac{\omega_p^2}{\omega^2}\cos^2\theta
    \end{matrix}\right)\cdot \left(\begin{matrix}
        E_x \\
        E_y \\
        E_z
    \end{matrix}\right)\, .
\end{equation}

\section{Axion electrodynamics} \label{app:axion-electrodynamics}
In this appendix, we revisit the derivation of axion-photon conversion probability\citep{2018PhRvL.121x1102H,2020PhRvD.102b3504B,2021PhRvD.104j3030W,2021JCAP...11..013M,1988PhRvD..37.1237R}, i.e., \EQ{eq:convprob}. The equation of motion of axion fields and electric fields is given by Eqs.~(\ref{eq:axion-EOM}) and (\ref{eq:E-EOM}) and is reshown here,
\begin{align}
    -\partial_t^2a +\nabla^2a - m_a^2 a &= -g_{a\gamma\gamma}{\mathbf{E}}\cdot {\mathbf{B}}\, , \label{eq:axion-EOM-app} \\
    -\nabla^2 {\mathbf{E}}+\nabla(\nabla\cdot {\mathbf{E}}) - \omega^2 {\mathbf{D}} &= \omega^2 g_{a\gamma\gamma}a{\mathbf{B}}\, . \label{eq:E-EOM-app}
\end{align}
One can regard both equations as the wave equations driven by source terms on the right-hand side.
We take the axion's direction of motion to be $\hat{z}$ and the direction of the magnetic field described below \EQ{eq:electric-displacement}. When considering the case of axion-photon mixing in a slowly varying and locally uniform plasma, all derivatives that do not involve the derivative in the $z$ direction can be neglected. Therefore, with the expression of electric displacement tensor \EQ{eq:electric-displacement}, the $x$ component of \EQ{eq:E-EOM-app} can be simplified to
\begin{equation}
    -\partial_z^2 E_x +\partial_{xz} E_z = \omega^2 E_x\, ,
\end{equation}
with the planar wave approximation, derivatives that involve the $x$ direction (or $y$ direction) can be neglected. Therefore, from the above equation, we can see that $E_x$ does not couple to the axion fields directly. Then the $y$ component of \EQ{eq:E-EOM-app},
\begin{align}
    \nonumber 0 =& \left(\omega^2-\omega_p^2\sin^2\theta\right)E_y + \omega_p^2\cos\theta\sin\theta E_z \\
    &+  \frac{\partial^2 E_y}{\partial z^2} + \omega^2 g_{a\gamma\gamma}aB\sin\theta \, ,\label{eq:dezdzy}
\end{align}
and similarly for the $z$ component,
\begin{align}
    \nonumber 0=& \left(\omega^2-\omega_p^2\sin^2\theta\right)E_z +\omega_p^2\cos\theta\sin\theta E_y \\
    &+ \omega^2g_{a\gamma\gamma}aB\cos\theta\, . \label{eq:deydzy}
\end{align}
Plugging \EQ{eq:deydzy} into \EQ{eq:dezdzy}, one gets
\begin{equation}
    -\frac{\partial^2 E_y}{\partial z^2} = \frac{\omega^2-\omega_p^2}{1-\frac{\omega_p^2}{\omega^2}\cos^2\theta}E_y + \frac{\omega^2\sin\theta}{1-\frac{\omega_p^2}{\omega^2}\cos^2\theta}g_{a\gamma\gamma}aB\, ,
    \label{eq:b6}
\end{equation}
which is the first equation in \EQ{eq:eom-ey-a}. With a slowly varying plasma, the electric fields and axion fields can be written as
\begin{equation}
    E_y\equiv \tilde{E}_y(z) e^{i\left(\omega t-kz\right)}\, ,\quad a\equiv \tilde{a}(z) e^{i\left(\omega t-kz\right)}\, .\label{eq:tharm}
\end{equation}
Take the WKB approximation, which assumes that $k \partial\tilde{E}_y/\partial z \gg\partial^2 \tilde{E}_y/\partial z^2$. 
Neglecting the generation of secondary axions from self-interaction of the electromagnetic field [at $O(g_{a\gamma\gamma}^2)$], \EQ{eq:axion-EOM-app} leads to the axion energy-momentum relation $k^2+m_a^2-\omega^2=0$. 
With \EQ{eq:tharm}, \EQ{eq:b6} is further simplified to
\begin{equation}
    i\frac{\partial \tilde{E}_y}{\partial z} = \frac{1}{2k}\left(m_a^2-\xi\omega_p^2\right)\tilde{E}_y + \frac{\omega^2\xi}{2k\sin\theta}g_{a\gamma\gamma}B\tilde{a}\, ,   \label{eq:eom-sch}
\end{equation}
where we have defined
\begin{equation}
    \xi = \frac{\sin^2\theta}{1-\frac{\omega_p^2}{\omega^2}\cos^2\theta}\, .
\end{equation}
Note that \EQ{eq:eom-sch} has a solution in integration form,
\begin{equation}
    i\tilde{E}_y = \int_0^z\mathrm{d}z' \, \frac{1}{2k}\frac{\omega^2\xi}{\sin\theta}g_{a\gamma\gamma}B\tilde{a} e^{-i\int_0^{z'} \mathrm{d}z''\, \frac{1}{2k}(m_a^2-\xi\omega_p^2)}\, .
\end{equation}
This solution can be generalized as $\int_0^t\mathrm{d}s\, g(s)e^{if(s)}$. The WKB approximation computes the integral by finding the stationary trajectory, namely, $f'(s_0)=0$. After Taylor expanding the phase and taking the limits of integration to infinity, stationary phase approximation gives
\begin{align}
    \nonumber \int_0^t\mathrm{d}s\,g(s)e^{if(s)}\approx& g(s_0)e^{if(s_0)}\int_{-\infty}^\infty\mathrm{d}s\, e^{\frac{if''(s_0)}{2}(s-s_0)^2}    \\
    =& g(s_0)e^{if(s_0)+{\rm sign}[f''(s_0)]i\pi/4}\sqrt{\frac{2\pi}{\left|f''(s_0)\right|}}\, .
\end{align}
With this approximation, one can then derive the analytical expression of axion-photon conversion probability \EQ{eq:convprob}, which is defined as the energy flux (derived from the stress-energy tensor) ratio between the photon and axion fields \citep{2018PhRvL.121x1102H,2020Leroy}, 
\begin{equation}
    P_{a\to\gamma} \equiv \frac{F_\gamma}{F_a} = \frac{1}{\omega^2} \frac{k_\gamma^2}{k_a^2}\left|\frac{E(z)}{a(0)}\right|^2\simeq\frac{1}{\omega^2}\left|\frac{E_y}{a}\right|^2\, ,
\end{equation}
where $F_a\propto k_a^2|a|^2$ is the energy flux of plane wave axion field $a$, and similarly $F_\gamma$ is the energy flux for the photon field. $k_a$ and $k_\gamma$ are the momentum of the axion and photon fields, respectively.
The stationary point condition [$f'(s_0)=0$] is the resonance condition,
\begin{equation}
    m_a^2 = \xi\omega_p^2=\frac{\omega_p^2\sin^2\theta}{1-\frac{\omega_p^2}{\omega^2}\cos^2\theta}\, ,
\end{equation}
which simplifies to
\begin{equation}
    \omega_p(\mathbf{x}_c)^2 = \frac{m_a^2 \omega^2}{\omega^2\sin^2\theta+m_a^2\cos^2\theta}\, .
\end{equation}
As a summary of this section, due to the WKB approximation, we regard that the axion is converted to the photon \emph{locally}. The position of conversion is determined by the resonance condition, and the amplitude of the radio wave is described by the conversion probability.

\section{Propagation of polarized radio waves} \label{app:polarization}
In this appendix, we will show that the propagation of polarized radio waves can be computed from either electrodynamics, Fermi-Walker transport (spin theory), or parallel transport (optical theory).

\subsection{Derivation of propagation law}
Starting with Maxwell's equations,
\begin{align}
    \nabla\cdot\mathbf{D} &= 4\pi\rho\, , \\
    \nabla\cdot\mathbf{B} &= 0\, , \\
    \nabla\times\mathbf{E}+\partial_t{\mathbf{B}} &= 0\, , \\
    \nabla\times\mathbf{H}-\partial_t{\mathbf{D}} &= 4\pi\mathbf{j}\, .
\end{align}
Using $\mathbf{D} = \epsilon\mathbf{E}$ and $\mathbf{B}=\mu\mathbf{H}$, one obtains the wave equation
\begin{equation}
    \nabla^2\mathbf{E}-\epsilon\mu\partial_t^2{\mathbf{E}}+\left(\nabla\ln\mu\right)\times\nabla\times\mathbf{E}+\nabla\left(\mathbf{E}\cdot\nabla\ln\epsilon\right)=0\, , \label{eq:wave_eq_E}
\end{equation}
and a similar equation for $\mathbf{H}$,
\begin{equation}
    \nabla^2\mathbf{H}-\epsilon\mu\partial_t^2{\mathbf{H}}+\left(\nabla\ln\epsilon\right)\times\nabla\times\mathbf{H}+\nabla\left(\mathbf{H}\cdot\nabla\ln\mu\right)=0\, .
\end{equation}
Following \citep{1999prop.book.....B}, for the time-harmonic field we have
\begin{equation}
    \mathbf{E}(\mathbf{r},t) = \mathbf{e}(\mathbf{r}) e^{ik_0\mathcal{S} (\mathbf{r})} e^{-i\omega t}\, ,\, \mathbf{H}(\mathbf{r},t) = \mathbf{h}(\mathbf{r}) e^{ik_0\mathcal{S} (\mathbf{r})} e^{-i\omega t}\, , \label{eq:general_form}
\end{equation}
where $\mathbf{e}(\mathbf{r})$ and $\mathbf{h}(\mathbf{r})$ are vector fields of position, $k_0$ is the vacuum wave vector, and $\mathcal{S} (\mathbf{r})$ is the optical path length. The refractive index is defined as $n=\sqrt{\epsilon\mu}$. With these expressions \EQ{eq:general_form}, Maxwell's equations become
\begin{align}
    \mathbf{e}\cdot\nabla\mathcal{S} &= 0\, , \\
    \mathbf{h}\cdot\nabla\mathcal{S} &= 0\, ,\\
    \mathbf{e}\times\nabla\mathcal{S}+\mu\mathbf{h} &= 0\, ,\\
    \mathbf{h}\times\nabla\mathcal{S}-\epsilon\mathbf{e} &= 0\, ,
\end{align}
where we neglect terms that are higher than the first order of $1/(ik_0)$, and the medium (plasma) responses are described by $\epsilon$ and $\mu$.

Substitute the first equation of \EQ{eq:general_form} into the wave equation \EQ{eq:wave_eq_E} and one obtains
\begin{equation}
    \mathbf{K}(\mathbf{e},\mathcal{S},n)+\frac{1}{ik_0}\mathbf{L}(\mathbf{e},\mathcal{S},n,\mu)+\frac{1}{\left(ik_0\right)^2}\mathbf{M}\left(\mathbf{e},\epsilon,\mu\right) = 0\, ,
\end{equation}
where
\begin{align}
    &\mathbf{K}(\mathbf{e},\mathcal{S},n) = \left[n^2-\left(\nabla\mathcal{S}\right)^2\right]\mathbf{e}\, ,\\
    &\mathbf{L}(\mathbf{e},\mathcal{S},n,\mu) = \left(\nabla\mathcal{S}\cdot\nabla\ln\mu-\nabla^2\mathcal{S}\right)\mathbf{e}-2\left(\mathbf{e}\cdot\nabla\ln n\right)\nabla\mathcal{S} \nonumber \\
    &\qquad\qquad\qquad  -2\left(\nabla\mathcal{S}\cdot\nabla\right)\mathbf{e}\, , \\
    &\mathbf{M}\left(\mathbf{e},\epsilon,\mu\right) = \nabla  \times\mathbf{e}\times\nabla\ln\mu-\nabla^2\mathbf{e}-\nabla\left(\mathbf{e}\cdot\nabla\ln\epsilon\right)\, .
\end{align}

For sufficiently large $k_0$, $\mathbf{K} = 0$ is required at the first order. This is the eikonal equation that $\nabla\mathcal{S}/n$ is a unit vector, i.e.,
\begin{equation}
    \frac{\nabla\mathcal{S}}{n} = \frac{\nabla\mathcal{S}}{|\nabla\mathcal{S}|}\, .
\end{equation}
For isotropic media, the direction of the light ray can be taken as the direction of the averaged Poynting vector, which is
\begin{equation}
    \langle\mathbf{S} \rangle  = \frac{1}{8\pi}\Re\left(\mathbf{E}_0\times\mathbf{H}_0^\ast\right)\, ,
\end{equation}
where $\mathbf{E}_0$ and $\mathbf{H}_0$ are the spatial dependence of electromagnetic waves. With \EQ{eq:general_form}, this turns into 
\begin{equation}
    \langle\mathbf{S} \rangle = \frac{1}{8\pi\mu}\left(\mathbf{e}\cdot\mathbf{e}^\ast\right)\nabla\mathcal{S}\, .
\end{equation}
Therefore, the ray equation can be written as 
\begin{equation}
    \frac{d\mathbf{r}}{ds} = \frac{\nabla\mathcal{S}}{n}\, .
\end{equation}

To the next order of $ik_0$,  $\mathbf{L} = 0$, using $\partial/\partial\tau = \nabla\mathcal{S}\cdot\nabla$, equations  $\mathbf{L} = 0$ become
\begin{equation}
    \frac{\partial\mathbf{e}}{\partial\tau}+\frac{1}{2}\left(\nabla^2\mathcal{S}-\frac{\partial\ln\mu}{\partial\tau}\right)\mathbf{e}+\left(\mathbf{e}\cdot\nabla\ln n\right)\nabla\mathcal{S} = 0\, . \label{eq:prop_amplitude}
\end{equation}
Multiplying this by $\mathbf{e}^\ast$ and adding its complex conjugate, one finds
\begin{equation}
    \frac{\partial}{\partial\tau}\left(\mathbf{e}\cdot\mathbf{e}^\ast\right)+\left(\nabla^2\mathcal{S}-\frac{\partial\ln\mu}{\partial\tau}\right)\mathbf{e}\cdot\mathbf{e}^\ast = 0\, . \label{eq:e_dot_e^star}
\end{equation}
The polarization base vector ($\hat{\mathbf{f}}$) is the normalized vector of electric field amplitude that
\begin{equation}
    \hat{\mathbf{f}} = \frac{\mathbf{e}}{\sqrt{\mathbf{e}\cdot\mathbf{e}^\ast}}\, .
\end{equation}
Substituting into \EQ{eq:prop_amplitude}, one obtains 
\begin{align}
    & \frac{d \hat{\mathbf{f}}}{d\tau}=n\frac{d\hat{\mathbf{f}}}{ds}=-\left(\hat{\mathbf{f}}\cdot\nabla\ln n\right)\nabla\mathcal{S}\, , \\
    & \implies \frac{d\hat{\mathbf{f}}}{ds} = -\left(\hat{\mathbf{f}}\cdot\nabla\ln n\right)\hat{\mathbf{k}}\, , \label{eq:trans_polari_optics}
\end{align}
where \EQ{eq:e_dot_e^star} has been used in deriving the above equation. $s$ is the geometric length along the light ray and $\hat{\mathbf{k}}=d\mathbf{r}/ds$.

From the eikonal equation, one arrives at
\begin{align}
    & \frac{1}{n}\nabla\frac{d\mathcal{S}}{ds}=\frac{1}{n}\frac{d}{ds}\nabla\mathcal{S}\, , \\
    & \implies \frac{1}{n}\nabla\left(\nabla\mathcal{S}\cdot\frac{d\mathbf{r}}{ds}\right) = \frac{1}{n}\frac{d}{ds}\nabla\mathcal{S}\, , \\
    & \implies \nabla\ln n = \frac{1}{n}\frac{d}{ds}\left(n\hat{\mathbf{k}}\right)\, , \\
    & \implies \hat{\mathbf{f}}\cdot\nabla\ln n = \hat{\mathbf{f}}\cdot\left(\frac{d\ln n}{ds}\hat{\mathbf{k}}+\frac{d\hat{\mathbf{k}}}{ds}\right) = \hat{\mathbf{f}}\cdot\frac{d\hat{\mathbf{k}}}{ds}\, ,
\end{align}
where the transverse wave condition $\hat{\mathbf{k}}\cdot\hat{\mathbf{f}}=0$ is used in deriving the above equation. Plugging into \EQ{eq:trans_polari_optics}, then
\begin{equation}
    \frac{d\hat{\mathbf{f}}}{ds} = -\hat{\mathbf{k}}\left(\hat{\mathbf{f}}\cdot\frac{d\hat{\mathbf{k}}}{ds}\right)\, ,
    \label{eq:polp}
\end{equation}
which is exactly \EQ{eq:propagation-of-polari} shown in the main text \citep{2017mcp..book.....T}. Note that, following a similar procedure, the propagation law for the direction of $\mathbf{h}$ is the same equation shown above.

\subsection{Derivation using Fermi-Walker transport}
Any smooth vector field $v^a$ is said to be Fermi-Walker transported along a curve if
\begin{equation}
    \frac{\mathrm{D}_{\mathrm{F}} v^a}{ds} = \frac{\mathrm{D}v^a}{ds} + \left(A^a Z^b-Z^a A^b\right)v_b = 0\, ,
\end{equation}
where $\mathrm{D}v^a/ds$ is the covariant derivative of vector $v^a$, $Z^a$ is the covariant velocity, $A^a$ is the covariant acceleration, and $s$ is the affine parameter for this curve. In three-dimensional space, $s$ is the distance along the ray; if we choose the covariant velocity to be $\hat{\mathbf{k}}$, then the acceleration is given by
\begin{equation}
    \mathbf{A} = k^b\nabla_b \hat{\mathbf{k}} = \frac{d\hat{\mathbf{k}}}{ds}\, .
\end{equation}
Therefore, Fermi-Walker transport requires that the propagation for unit polarization vector $\hat{\mathbf{f}}$ is 
\begin{equation}
    \frac{d\hat{\mathbf{f}}}{ds}-\left(\frac{d\hat{\mathbf{k}}}{ds} \hat{\mathbf{k}}\cdot\hat{\mathbf{f}} - \hat{\mathbf{k}}\frac{d\hat{\mathbf{k}}}{ds}\cdot\hat{\mathbf{f}}\right) = 0\, .
\end{equation} 
The transverse wave condition requires $\hat{\mathbf{f}}\cdot \hat{\mathbf{k}}=0$, and the above equation simplifies to \EQ{eq:polp}.

\subsection{Derivation using parallel transport}
Equation (\ref{eq:propagation-of-polari}) or (\ref{eq:trans_polari_optics}) is equivalent to the parallel transport of the polarization vector in a curved space \citep{1964mtop.book.....L}.
Because of the refractive index, the effective line element for light propagation is
\begin{equation}
    d\sigma^2 = n^2ds^2  = n^2 (dx^2+dy^2+dz^2)\, ,
\end{equation}
i.e., the corresponding metric is
\begin{equation}
    g_{ij} = n^2(x,y,z)\delta_{ij}\, ,
\end{equation}
and the connections are
\begin{equation}
    \Gamma_{ij}^{k} = \frac{1}{n}\left(\delta_j^k\partial_i n+\delta_i^k\partial_j n-\delta_{ij}\partial^k n\right)\, .
\end{equation}

The parallel transportation of a vector field $v^a(\sigma)$ along a given curve $x^a(\sigma)$ is described by the covariant derivative that
\begin{equation}
    \frac{dv^a}{d\sigma}+\Gamma^a_{bc}v^b\frac{dx^c}{d\sigma} = 0\, .
\end{equation}
Plugging into the connection components, one obtains
\begin{equation}
    \frac{dv^k}{d\sigma}+\frac{1}{n}\partial_i nv^i\frac{dx^k}{d\sigma} + \frac{1}{n}\partial_j n v^k\frac{dx^j}{d\sigma} - \frac{1}{n}\partial^k n v_i \frac{dx^i}{d\sigma} = 0\, ,
\end{equation}
which can be rewritten as
\begin{equation}
    \frac{d\left(nv^k\right)}{d\sigma}+\frac{1}{n}\frac{dx^k}{d\sigma}\left(n v^i\cdot\partial_i n\right)-\frac{1}{n}\partial^k n\left(nv_i\cdot\frac{dx^i}{d\sigma}\right) = 0\, . \label{eq:parallel_transport_vec}
\end{equation}
Parallel transport of the tangential vectors $\frac{dx^i}{d\sigma}$ yields a similar expression
\begin{align}
    \frac{d}{d\sigma}\left(n\frac{dx^k}{d\sigma}\right) &+ \frac{1}{n}\left(n\frac{dx^i}{d\sigma}\cdot\partial_i n\right)\frac{dx^k}{d\sigma}\nonumber \\
    &-\frac{\partial^k n}{n}\left(n\frac{dx_i}{d\sigma}\cdot\frac{dx^i}{d\sigma}\right) = 0\, .
\end{align}
The two equations shown above can be simplified to
\begin{equation}
    \frac{d}{d\sigma}\left(n^2 v_i \cdot\frac{dx^i}{d\sigma}\right) = 0\, .
\end{equation}
Therefore, if $v^a$ is perpendicular to $\frac{dx^a}{d\sigma}$ at one point on the curve, then this holds at all points on the curve. Thus, the third term of \EQ{eq:parallel_transport_vec} vanishes,
\begin{equation}
    \frac{d}{d\sigma}\left(nv^k\right)+\frac{1}{n}\left(n v^i\cdot\partial_i n\right)\frac{dx^k}{d\sigma} = 0\, .
\end{equation}
Comparing this with \EQ{eq:trans_polari_optics}, one finds that $\frac{1}{n}\mathbf{f}$ satisfies the above condition and hence proves the parallelism of the directions $\hat{\mathbf{f}}$ along the curve.

\section{Convergence and robustness tests}    \label{app:convergence}
In this appendix, we describe different convergence checks and robustness tests of our codes. These include the tests of integral and model dependence.

\subsection{Integrals}
\begin{figure*}
    \centering
    \includegraphics[width=0.8\hsize]{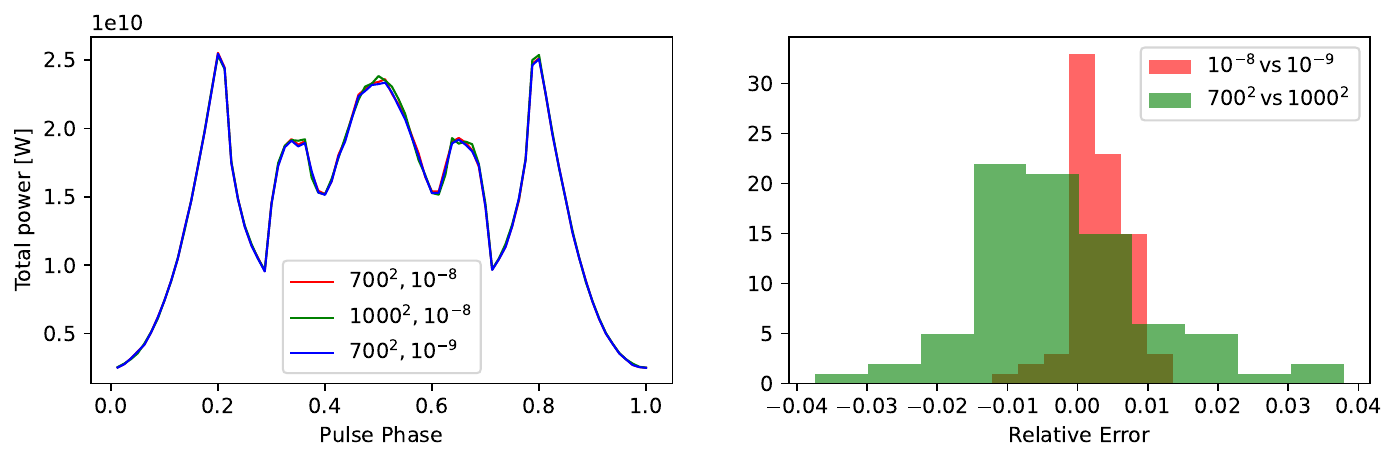}
    \caption{Convergence checks. Left: the comparison of axion-induced emission pulse profiles with different choices of number of pixels ($N=700^2,\, 1000^2$) and different maximum step sizes and error tolerance in integrating the photon trajectories [$h = r/10^4, 10^{-8}; \,h= r/(3\times 10^4), 10^{-9}$]. Right: histogram of relative error of pulse profile of each pulse phase. Note that the relative error of pulse profiles is different from the error tolerance of photon trajectory.
    The green bars are for the difference between the case of $N=700^2$ and $1000^2$, whereas the red bars are for the difference between photon trajectory tolerance of $h=r/10^4,10^{-8}$ and $h=r/(3\times 10^4),10^{-9}$.}
    \label{fig:convergence}
\end{figure*}

The integrals in our calculation can be divided into two parts. One is the integral on the projection image to obtain the total axion-induced emission power and polarization. The other one is the integral of geodesic \EQ{eq:geodesic} to obtain the photon's trajectory. To verify the convergence of these integrals, we try to compare the result with a larger number of pixels and a lower relative error tolerance. In the left panel of Fig.~\ref{fig:convergence}, we show the axion-induced emission pulse profiles of viewing angle $36^\circ$ and axion mass $m_a=0.5\ {\rm \mu eV}$ with different numbers of pixels ($N=700^2,\, 1000^2$) and different choices of adaptive size control. One has a maximum step size of $r/10^4$ with relative error tolerance $10^{-8}$; the other one has a maximum step size of $r/(3\times 10^4)$ with relative error tolerance $10^{-9}$. The right panel is the relative error of these different choices. 

\subsection{Magnetic field configuration}
In the main text, we assume an exact rotating dipole field configuration for a compact star's magnetic fields, which is unlikely in reality. So, this code test is to find out the consequence if we add a perturbation to the dipole field configuration.

\begin{figure}
    \centering
    \includegraphics[width=0.9\hsize]{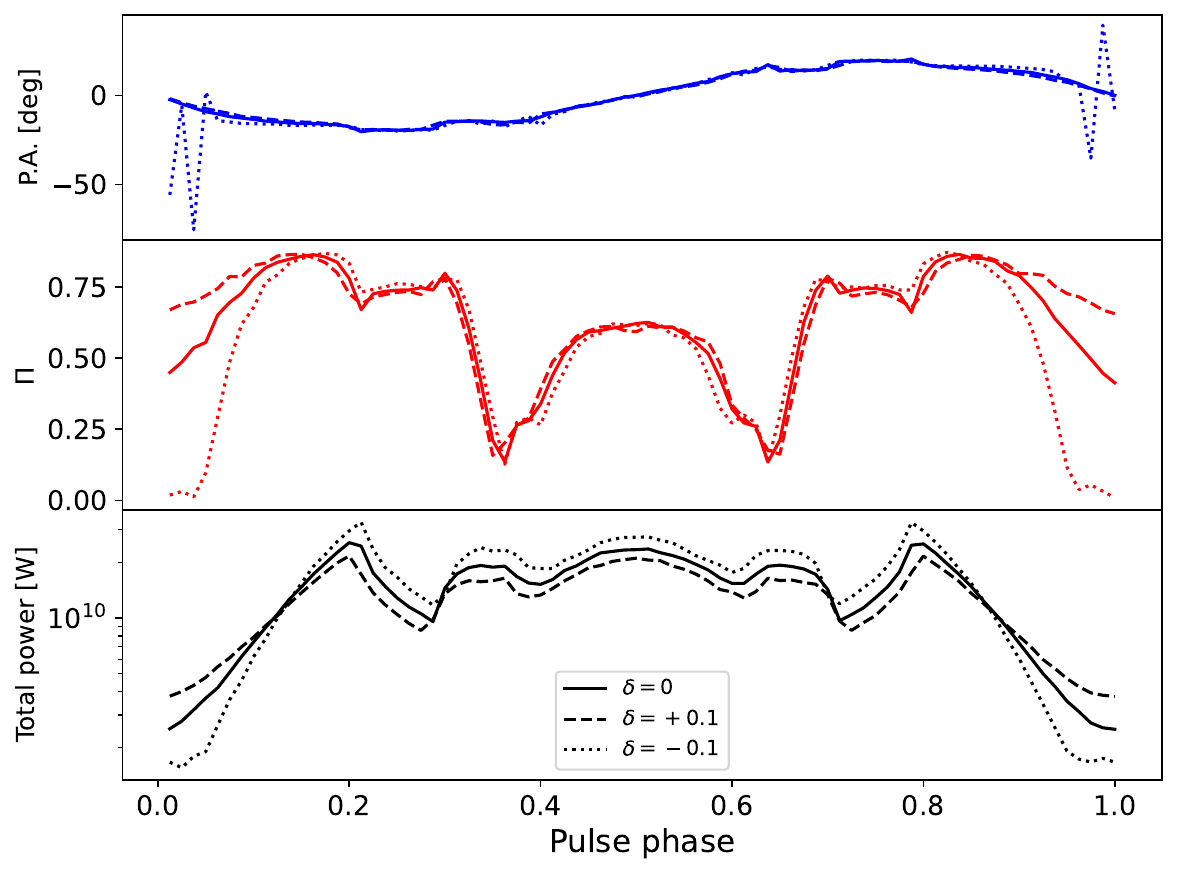}
    \caption{Robustness tests. A comparison of axion-induced emission considering a perturbation to the dipole magnetic field configurations where the exponent of $r$ dependence is $3+\delta$.}
    \label{fig:conv-dipole}
\end{figure}

First, we assume the exponent of $r$ dependence deviates from 3; therefore, the magnetic field turns to
\begin{align}
    B_r &= B_0\left(\frac{R}{r}\right)^{3+\delta}\left(\cos\chi\cos\theta+\sin\chi\sin\theta\cos\psi\right)\, ,     \\
    B_\theta &= \frac{B_0}{2}\left(\frac{R}{r}\right)^{3+\delta}\left(\cos\chi\sin\theta-\sin\chi\cos\theta\cos\psi\right)\, ,    \\
    B_\phi &= \frac{B_0}{2}\left(\frac{R}{r}\right)^{3+\delta}\sin\chi\sin\psi\, .
\end{align}
For a slight deviation, we choose $\delta=0.1,\, -0.1$, and the result is shown in Fig.~\ref{fig:conv-dipole}.

\begin{figure}
    \centering
    \includegraphics[width=0.9\hsize]{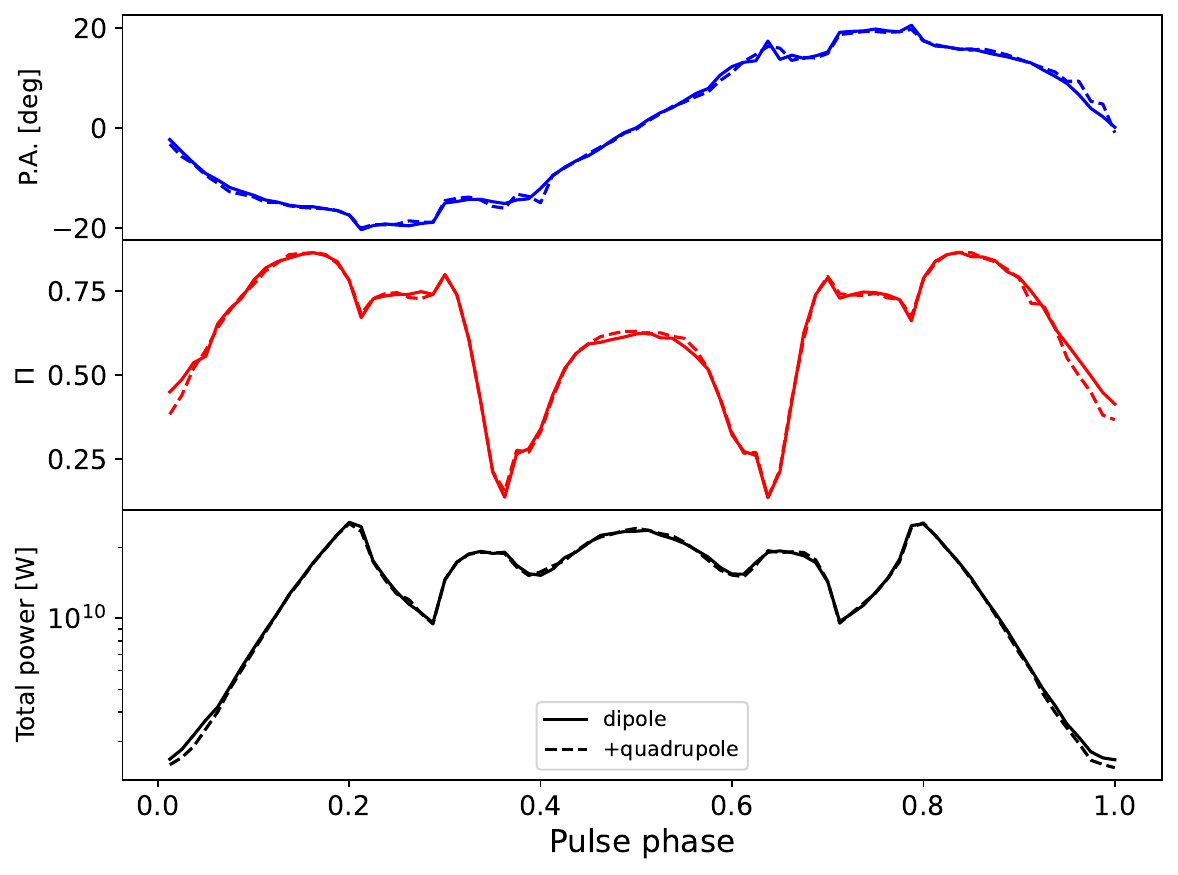}
    \caption{Robustness tests.  A comparison of axion-induced emission considering a perturbation to the dipole magnetic field configuration where a quadrupole component is included to perturb the surface magnetic field by approximately 10\%.}
    \label{fig:conv-quad}
\end{figure}

Second, we assume a perturbation to the dipole magnetic field. Then, the total magnetic field can be written as
\begin{equation}
    \mathbf{B} = \mathbf{B}^D+\mathbf{B}^Q\, ,
\end{equation}
where $\mathbf{B}^D$ is given by Eqs.~(\ref{eq:dipolefield1}--\ref{eq:dipolefield3})
and a quadrupole magnetic field component is given by
\begin{align}
    B_r^Q &= \frac{B_0\epsilon}{2}\left(\frac{R}{r}\right)^{4}\left(3\cos2\theta+1\right)\, ,     \\
    B_\theta^Q &= B_0\epsilon\left(\frac{R}{r}\right)^{4}\sin2\theta\, ,    \\
    B_\phi^Q &= 0\, .
\end{align}
For $\epsilon$, we choose $\epsilon=0.1$, and
 the result of this perturbation is shown in Fig.~\ref{fig:conv-quad}.

\bibliography{ref}

\end{document}